\documentstyle[a4,aps,preprint]{revtex}

\title{LATTICE BOLTZMANN MODEL FOR MAGNETIC FLUIDS}
\author{Victor SOFONEA}
\address{Research Center for Hydrodynamics, Cavitation and
Magnetic Fluids\\ Technical University of Timi\c soara\\
Bd. Mihai Viteazul 1, R-1900 Timi\c soara, Romania}

\begin{document}

\maketitle

\begin{abstract}
A lattice Boltzmann model with interacting particles was developed in order to
simulate the magneto-rheological characteristics of magnetic fluids. In the
frame of this model, $6\, +\,1$ species of particles are allowed to move across
a $2D$ triangular lattice. Among these species, $6$ of them carry an individual
magnetic dipole moment and interact themselves not only as a result of
local collisions, as in current Lattice Boltzmann models, but also as a result
of nearest neighbours magnetic dipole-dipole interaction. The relative
distribution of the individual magnetic moments is determined by the intensity
of an external static magnetic field acting on the whole system.

This model exhibits some relevant characteristics of real magnetic fluids,
i.e.,
anisotropic structure formation as a result of magnetic field induced
gas-liquid phase transition and magnetic field dependence of the sound
velocity and the attenuation coefficient.
\end{abstract}
\begin{center}
{\bf {Keywords}}

magnetic fluids; computational techniques; structure formation
\end{center}

\newpage
\begin{section}{Introduction}
Magnetic fluids, also known as ferrofluids, are ultrastable colloidal
suspensions of subdomain ferro - or ferrimagnetic particles - e.g.,
magnetite $(Fe_3O_4)$ - dispersed in various carrier liquids (e.g.,
water, petroleum, transformer oil, organic solvents, alcohols). These
suspensions behave like quasihomogeneous strongly
magnetizable liquids due to the presence of approx.
$10^{17}\,-\,10^{18}$ magnetic particles in one cubic centimeter and
unite the properties of magnetic materials with those of fluids in a
rather spectacular way.

Many experimental results suggested that colloidal particles in
magnetic fluids always coagulate and form chain clusters, this process
being enhanced in the presence of a magnetic field. The formation of
chain clusters was observed with an electron microscope \cite{b2}. The
chain formation process, together with the reorientation of individual
particles in the presence of a magnetic field, are responsible for the
anisotropy of the physical properties of the magnetic fluids
\cite{b3}. For example, the transversal magneto-optical effects
(birefringence and linear dichroism) induced in thin magnetic fluid
layers are well explained by the above-mentioned microstructural
processes \cite{b4,b5}. The sound velocity and the acoustic attenuation
coefficient in magnetic fluids are also depending on the angle between
the sound propagation direction and the external magnetic field
\cite{b6}.

Several theoretical works \cite{b3,b10,b11,b12} and even computer
simulations \cite{b13,b14,b15,b16,b17} were dedied to particle
interactions in magnetic fluids. The Monte Carlo \cite{b13,b14,b15},
as well as other computer simulations \cite{b16,b17} performed
up to date were limited to magnetic fluids considered only as
magnetizable media, but the fluid behaviour of these materials were
not considered using such techniques.

The coming out of lattice gas models gave the
possibility to introduce particle interactions (other than collisions)
in the frame of simple
models \cite{b19,b20,b21,b22} e.g., for immiscible fluids. The
recent established Lattice Boltzmann methods having the capability to
consider not only the rheological behaviour of multiple components
fluids, but also their long range particle interactions \cite{b23,b24,b25},
encouraged us to develop the subsequent appropriate model for magnetic fluids
\cite{x1,x2}.
\end{section}
\begin{section}{Lattice Boltzmann model}
\begin{subsection}{Lattice description}

It is known that  fluid phenomena seen in nature are the statistical
behaviours of their associated microsystems. The rheological and the
thermodynamical properties of a fluid system are, therefore, natural
consequences of the dynamics of these physical microsystems. Due to
the nonunique correspondence between a fluid system and a microsystem,
artificial microsystems may be constructed, which are simple enough to
be simulated on a computer, but yet contain all the required physics
of a realistic fluid system.

Lattice gas models, whose first one was the FHP modfel \cite{b26},
simulate the exact dynamical history of an integer number of particles
moving on a regular lattice while conserving mass and momentum during
their collisions. The single-particle equilibrium distribution
function specifies the fluid density at each lattice site and also the
velocity state, while equilibrium is determined by the particle
collision rules. The current trend in cellular automata fluid
simulations is to replace the lattice-gas (LG) approach by the Lattice
Boltzmann (LB) methods \cite{b23,b24,b25,b27,b28,b29,b30}. A Lattice
Boltzmann automaton uses a real-number description for the particle
distribution and is far less noisy than the LG approach. It is
parallel in nature, due to the fact that all the information transfer
is local in time and space, so that it is most suitable for the
massively parallel computers.

Following the general LB approach \cite{b25}, the following lattice
Boltzmann equations were considered for a fluid with $S+1$ total components on
a two-dimensional ($2D$) hexagonal lattice:
\begin{eqnarray}\label{eboltz}
n^\sigma_a(\roarrow{x}+\hat e_a,t+1)\,-\,n^\sigma_a(\roarrow{x},t) & = &
 \Omega^\sigma_a(\roarrow{x},t)\\
\sigma\,=\,0,1,\ldots ,S & &\nonumber\\
a\,=\,0,1,\ldots ,b & &\nonumber
\end{eqnarray}
where $n_a^\sigma(\vec x,t)$ is the single particle distribution function for
 the $\sigma$-th component
 having the velocity directed along the vector $\hat e_a$ (see below) and
 $\Omega_a^\sigma (\vec x,t)$ is the collision term. In order to simplify the
 computer program as much as possible, while retaining the relevant physical
 aspects, in our investigations we considered that all particles have the same
 mass, which
 is set equal to $M^\sigma = 1$ for all $\sigma = 0,1,\ldots ,S$.

The particles are located at the nodes of the $2D$ triangular lattice generated
 by the unit vectors $\roarrow{e_1^0}\,=\,(\,1\, ,\,0\,)$ and
$\roarrow{e_2^0}\,=\,(\,\frac{1}{2}\, ,\,\frac{\sqrt 3}{2}\,)$ so that any
particle position vector $\vec x\,$ is a linear combination of these
 generating vectors. Since all our investigations were performed on simple
 geometry $2D$ lattices having almost a rectangular
form, the nodes were numbered from $0$ to $ndx$ in the $X$ direction and from
 $0$ to $ndy$ in the $Y$ direction. Consequently, any particle position vector
$\vec x_{ij}$ is determined by its corresponding pair of indices $(i,j)$. All
the subsequent computer simulations were made with $ndx = ndy =127$ under
periodic boundary conditions.

At each time step, particles can be at rest or can move to their nearest
neighbour
nodes. The nearest neighbours of any site $\roarrow{x}$ are
 $\roarrow{x}_a\,=\,\roarrow{x}\,+\,\hat e_a$ \cite{b31},
 where $|\hat e_a|\,=\,c\,=1$ (the lattice constant) and
\begin{equation}
\frac{1}{c}\,\hat e_a\,=\,\left(\,(\hat e_a)_1\, , (\hat e_a)_2\,\right)\,
=\,\left(\,\cos\frac{2\pi(a-1)}{b}\,,\,\sin\frac{2\pi(a-1)}{b}\,\right)
\end{equation}
for all $a\,=\,1,\ldots,b$ ($b\,=\,6$ for the triangular lattice).
For convenience, as usual in LB methods, the vector $\hat e_a, \,a=1,\ldots b$
has also the meaning of the velocity of a particle moving in the $a$ direction
while the vector $\hat e_0,\,|\hat e_0|=0$ is associated vith the velocity of
rest particles.

For simplicity, we adopted a single relaxation time form for the collision
 term and so,
\begin{equation}
\Omega_a^\sigma (\roarrow{x},t)\,=\,-\,\frac{1}{\tau}\,[\,n_a^\sigma
 (\roarrow{x},t)
\, - \,n_a^{\sigma,eq} (\vec x,t)\,]
\end{equation}
where $\tau$ is the mean collision time
and $n_a^{\sigma,eq}$ is the equilibrium distribution with a given functional
form at site $\vec x$ and time $t$. The following form for $n_a^{\sigma,eq}$
is adopted from \cite{b23,b24,b25}:
\begin{eqnarray}\label{nequi}
n_a^{\sigma,eq} & = & n^\sigma\left[\,\frac{1-d_0}{b}\,+\,\frac{D}{c^2b}\,
(\hat e_a\cdot \vec u^\sigma)\,+\,\frac{D(D+2)}{2c^4b}\,(\hat e_a\cdot \vec
u^\sigma)^2\,
-\,\frac{D}{2c^2b}\,(\vec u^\sigma\cdot\vec u^\sigma)\,\right]\nonumber
\medskip\\
n_0^{\sigma,eq} & = & n^\sigma\,\left[\,d_0\,-\,\frac{1}{c^2}\, (\vec
u^\sigma\cdot \vec u^\sigma)\right]
\end{eqnarray}
where
\begin{equation}\label{eu}
n^\sigma\,=\,\sum_{a=0}^{a=b}\,n_a^\sigma\qquad and\qquad
\vec u^\sigma\,=\,\frac{1}{n^\sigma}\,\sum_{a=0}^{a=b}\,n_a^\sigma\hat e_a
\end{equation}
are respectively the number density and the averaged particle velocity for the
$\sigma$-th component at any $(\vec x,t)$ after collision and $d_0\,\le\,1$ is
 a constant. Moreover, when equilibrium is reached, one has $\vec
u^\sigma (\vec x,t)\, =\,\vec u (\vec x,t)$ for all $\sigma = 0,\ldots S$ and
\begin{equation}\label{utot}
\vec u (\vec x,t)\,\sum_{\sigma=0}^S\,n^\sigma(\vec x,t)\, =\,
\sum_{\sigma=0}^S\sum_{a=1}^b
\hat e_a n_a^\sigma(\vec x,t)
\end{equation}
For sufficiently small $|\vec u^\sigma |$, the above forms for the equilibrium
distribution functions will be positive.
\end{subsection}

\begin{subsection}{Interaction potential}
In order to apply the general LB model to magnetic fluids, we consider the
particles corresponding to $\sigma=0$ as being the ``carrier liquid''
particles,
while the ``colloidal particles'' are those corresponding to $1\leq\sigma\leq S
$. The colloidal particles carry a magnetic moment of magnitude $m\,=\,1$,
whose
orientation is fixed during the simulation process. For simplicity, we adopted
$S=6$, so that the magnetic moments of each kind of colloidal particles are the
vectors:
\begin{equation}
\vec m^\sigma\,=\,\left(\,m_1^\sigma\, ,\,m_2^\sigma\,\right)\, = \,
m\,\left(\,\cos\frac{2\pi(\sigma-1)}{S}\, ,\,
\cos\frac{2\pi(\sigma-1)}{S}\,\right)
\end{equation}

The local density of particles is:
\begin{equation}\label{roloc}
\rho(i,j,t)\, =\,\sum_{\sigma=0}^{S}\,n^\sigma(i,j,t)
\end{equation}
while the averaged one is:
\begin{equation}
\bar\rho(t)\, \equiv\, <\rho(t)> \, =
\,\frac{1}{(ndx+1)(ndy+1)}\,
\,\sum^{ndx}_{i=0}\sum^{ndy}_{j=0}\sum^{S}_{\sigma =0}\,
n^\sigma (i,j,t)
\end{equation}
The ``concentration'' of the colloidal particles is defined by
\begin{equation}
\phi\,=\,\frac{1}{<\rho >}<\sum_{\sigma=1}^{S}\, n^\sigma>
\end{equation}
and the local velocity $\vec u(\vec x,t)$ of the fluid is defined by
Eq.(\ref{utot}).

The solid particles in magnetic fluids are analogous to the molecules of a
 paramagnetic gas.
In the presence of an applied field and at a given absolute
 temperature $T$, the probability of finding a colloidal particle with a
given orientation $\sigma\,=\,1,\ldots ,S$ becomes proportional to the
Boltzmann factor
$\exp(-W^\sigma/kT)$, where $W^\sigma$ is the potential energy of the $\sigma$
-orientation magnetic particles:
\begin{equation}
\frac{W^\sigma}{kT}\,=\,-\,\frac{\mu_0\,\vec m^\sigma\cdot \vec H}{kT}\,=\,
-\,h\,(\,\hat e_\sigma\cdot\vec H\,/\,H\,)
\end{equation}
with $h=\mu_0mH/kT$, $\mu_0$ being the magnetic permittivity of the vacuum.

{}From the above mentioned considerations, one can see that the parameters
$\bar\rho,\,\phi$ and $h$ are the main characteristics for any computer run
on the basis of our LB model. In order to study the structure formation, we
 always started our computer runs by assuming
that colloidal particles are initially quasi-homogeneously distributed over
the lattice
with a small $1\%$ random perturbation. Consequently, at $t=0$, we always had
\begin{eqnarray}\label{e319}
n^\sigma (i,j,t=0) & = &
\bar\rho\phi f^\sigma(\vec H)\,
\left(1+\frac{rand(i,j)-0.5}{100}\right)\nonumber\\
& &(\sigma\, =\, 1,\ldots ,S)\nonumber\\
n^0(i,j,t=0) & = & \bar\rho\, -\,\sum_{\sigma=1}^{S}n^\sigma(i,j,t=0)
\end{eqnarray}
where $0\leq rand(i,j)\le 1$ are uniformly distributed random numbers, $
0\leq i\leq ndx\, ,\, 0\leq j\leq ndy$ and
\begin{equation}\label{fsigma}
f^\sigma (\vec H)\, =\,\frac{\exp(\mu_0\,\vec m^\sigma\cdot\vec H/kT)}
{\sum_{\sigma=1}^{\sigma=S}\exp(\mu_0\,\vec m^\sigma\cdot\vec H/kT)}
\end{equation}
 At the beginning of each run, the magnetic fluid was always considered to be
 at rest, so that
\begin{equation}
\vec u^\sigma(i,j,t=0)\, =\,0
\end{equation}
This implies
\begin{eqnarray}\label{e322}
n_a^\sigma(i,j,t=0) & = & \frac{1-d_0}{b}\, n^\sigma(i,j,t=0)\, ,\,
 a=1,\ldots,b\medskip\nonumber\\
n_0^\sigma(i,j,t=0) & = & d_0\, n^\sigma(i,j,t=0)
\end{eqnarray}

During the time evolution, magnetic colloidal particles are supposed to
interact themselves via dipole-dipole interactions, which are not
velocity-dependent.
The other kinds of interactions, e.g., Van der Waals forces and surfactant
interaction is not considered here for the sake of simplicity. Consequently,
there are no interactions between two particles when almost one has
$\sigma\,= \,0$. Furthermore, only nearest-neighbours interactions are taken
into account
as in \cite{b25} and so, the total potential energy
at each site is, at any time $t$ \cite{b25}:
\begin{equation}
W(\vec x)\, =\, \sum_{\sigma=0}^{\sigma=S}\,n^\sigma(\vec x)\,V^\sigma(\vec x)
\end{equation}
with the specific potential density
\begin{equation}
V^\sigma(\vec x)\, =\, \sum_{\bar\sigma =0}^{\bar\sigma =S}\sum_{a=1}^{a=b}\,
G_{\sigma\bar\sigma a}n^{\bar\sigma}(\vec x+\hat e_a)
\end{equation}
The magnitude of $G_{\sigma\bar\sigma a}$ is, according to the general
expression \cite{b1} of the interaction energy of two magnetic dipoles:
\begin{equation}\label{ega}
G_{\sigma\bar\sigma a}\, =\,\frac{1}{c^3}\left[\,\vec m^\sigma\cdot\vec m^{
\bar\sigma}\,-\,\frac{3(\vec m^\sigma\cdot\hat e_a)(\vec m^{\bar\sigma}\cdot
\hat e_a)}{c^2}\right]
\end{equation}
which becomes simplified if one takes $c=1$, as mentioned above. Since neutral
``carrier liquid'' particles do not interact at all, we always have
$G_{\sigma\bar\sigma a}=0$ for any $a=1,\ldots ,6$ when $\sigma =0$ or
$\bar\sigma =0$.

Having the interaction potential defined, the rate of net momentum
change induced at each site is a simple generalisation of an
expression
in \cite{b25}:
\begin{equation}\label{demom}
\frac{d\vec u^\sigma(\vec x,t)}{dt}\, =\, -\, n^\sigma(\vec x,t)
\sum_{\bar\sigma=1}^{\bar\sigma=S}\sum_{a=1}^{a=b}\,G_{\sigma\bar\sigma a}
n^\sigma(\vec x+\hat e_a,t)\,\hat e_a
\end{equation}
(the fact that all particles have the mass equal to 1 has been taken into
 account).

The interaction process is achieved during the collision phase in the LB
automaton, i.e., during the collision time $\tau_\sigma\equiv\tau$. The effect
of the interaction is to modify the local velocities. Therefore, after the
interaction
is achieved, the new net momentum $\vec u_{new}^\sigma(\vec x,t)$ at site
$\vec x$ for the $\sigma$-th component becomes
\begin{equation}\label{eta}
\vec u_{new}^\sigma(\vec x,t)\, =\,\frac{1}{n^\sigma(\vec x,t)}\left[
n^\sigma(\vec x,t)\vec u^\sigma_{old}(\vec x,t)\, +\,\tau\frac{d\vec u^\sigma(
\vec x,t)}{dt}\,\right]
\end{equation}
where $\vec u^\sigma_{old}(\vec x,t)$ is the local velocity before the
interaction. In completely uniform equilibrium, there can be no
relative flow of particles of different species since these are
supposed to have the same mass. Consequently, the particle
distribution functions must be locally proportional \cite{b32}.
Otherwise, different kinds of particles would have different
temperature, which is unphysical. For this reason, the local velocity
$\vec u^\sigma_{old}(\vec x,t)$ before the interaction should be choosen
always as being the same for all
$\sigma$, i.e., one has
\begin{equation}
\vec u^\sigma_{old}(\vec x,t)\, =\,\vec u(\vec x,t)
\end{equation}
where $\vec u(\vec x,t)$ was defined by Eq.(\ref{utot}).

The interaction potential does not conserve the net momentum at each site, as
usual in current LG and LB methods. However, the total net momentum is
conserved on the whole lattice. This can be seen from the symmetry properties
of
$G_{\sigma\bar\sigma a}$:
\begin{eqnarray}
G_{\sigma\bar\sigma a} & = & G_{\bar\sigma\sigma a}\\
G_{\sigma\bar\sigma a} & = & G_{\sigma\bar\sigma mod(a+3,b)}
\end{eqnarray}
($1\leq a\leq b$ and $0\leq \sigma ,\bar\sigma\leq S$) since the term
$\tau\frac{d\vec u^\sigma(\vec x,t)}{dt}$ in Eq.(\ref{eta}) cancels when
summing over all $\vec x\equiv\vec x_{ij}$ in the lattice.
\end{subsection}
\begin{subsection}{Summary of the automaton rules}
The succesive operations to be performed at each lattice node at each time
step in the frame of our Lattice Boltzmann automaton are resumed as follows:
\begin{enumerate}
\item Given $n^\sigma_a(i,j,t)$, the value of $\vec u(i,j,t)$ is computed
according to Eq.(\ref{utot}).
\item The rate of net local momentum change for all kinds of particles,
induced at each site is evaluated according to Eq.(\ref{demom}).
\item The new velocities $\vec u_{new}^\sigma(i,j,t)$ are evaluated in
accordance with Eq.(\ref{eta}), where $\vec u_{old}^\sigma(i,j,t)\,
=\,\vec u(i,j,t)$.
\item The new equilibrium distribution functions $n_a^\sigma(i,j,t)$ are
computed in accordance with Eqs.(\ref{nequi}) where $\vec u^\sigma(i,j,t)\,
=\,\vec u^\sigma_{new}(i,j,t)$.
\item The Boltzmann equation Eq.(\ref{eboltz}) and the propagation
step are now considered in order to get the propagated distribution functions
$n_a^\sigma
(\vec x +\vec e_a,t+1)$.
\end{enumerate}
\end{subsection}
\begin{subsection}{Conservation laws}
The lattice Boltzmann equation
\begin{equation}\label{lbe}
n_a^\sigma(\vec x+\hat e_a,t+1)\, -\, n_a^\sigma(\vec x,t)\, = \,-\,\frac
{1}{\tau}\,(n_a^\sigma(\vec x,t)\, -\, n_a^{\sigma,eq})
\end{equation}
can be rewritten after performing a Taylor's expansion ($n_a^\sigma\equiv
n_a^\sigma(\vec x,t)$; summation from $1$ to $2$ over repeated greek indices
 is understood):
\begin{equation}\label{lbes}
\partial_t n_a^\sigma\, +\,(\hat e_a)_\beta\partial_\beta
n_a^\sigma\,+\,\frac{1}{2}\,(\hat e_a)_\gamma(\hat e_a)_\beta\partial_\gamma
\partial_\beta n_a^\sigma\, =\,-\,\frac{1}{\tau}\,(n_a^\sigma\, -\,
n_a^{\sigma,eq})
\end{equation}
Retaining only the first-order derivatives and summing over $\sigma$ and $a$,
 one has \cite{b31}:
\begin{equation}
\partial_t\sum_{\sigma,a}\,n_a^\sigma\,
+\,\sum_{\sigma,a}\,\partial_\alpha
(\hat e_a)_\alpha n_a^\sigma\, =\,0
\end{equation}
which, according to Eqs. (\ref{utot}) and (\ref{roloc}), is just the continuity
(mass) equation:
\begin{equation}\label{econt}
\partial_t\rho\, +\,\nabla\cdot(\rho\vec u)\, =\, 0
\end{equation}

The momentum conservation equation is obtained when Eq.(\ref{lbes}) is
multiplied by $(\hat e_a)_\alpha$ and summed over $\sigma$ and $a$; th
e supplementary term
 $-\partial_\alpha W$ was added because of the interaction potential:
\begin{equation}\label{eqm}
\partial_t(\rho u_\alpha)\, +\, \partial_\beta\Pi_{\alpha\beta}\, +\,
P_\alpha\, =\,-\,\frac{1}{\tau}\,\sum_{\sigma,a}\,(\hat e_a)_\alpha
n_a^{\sigma,
neq}\, -\,\partial_\alpha W
\end{equation}
where
\begin{eqnarray}
\rho  u_\alpha & = & \sum_{\sigma,a}\,(\hat e_a)_\alpha\,n^\sigma_a\\
\Pi_{\alpha\beta} & = & \sum_{\sigma,a}\,(\hat e_a)_\alpha(\hat e_a)_\beta
n_a^\sigma\\
P_\alpha & = & \frac{1}{2}\,\partial_\beta\partial_\gamma\sum_{\sigma,a}
\,(\hat e_a)_\alpha(\hat e_a)_\beta(\hat e_a)_\gamma n_a^\sigma\\
n_a^{\sigma,neq} & = & n_a^\sigma\, -\, n_a^{\sigma,eq}\\
-\,\partial_\alpha\,W & = & \sum_\sigma\,\frac{du_\alpha^\sigma}{dt}
\, = \,
-\,\sum_{\sigma ,\bar\sigma ,a}\,G_{\sigma\bar\sigma a}n^\sigma(\vec x)
n^{\bar\sigma}(\vec x+\hat e_a)(\hat e_a)_\alpha
\end{eqnarray}
According to \cite{b31}, the term $\sum_{\sigma,a}(\hat e_a)_\alpha
 n_a^{\sigma,neq}$ vanishes.

Retaining only the first-order derivatives in the LB equation (\ref{lbes}) and
assuming
\begin{eqnarray}
|\,n_a^{\sigma,neq}\,| & \ll & |\, n_a^{\sigma,eq}\,|\\
|\,\partial_t n_a^{\sigma,neq}\,| & \ll & |\,\partial_t n_a^{\sigma,eq}\,|\\
|\,\partial_\gamma n_a^{\sigma,neq}\,| & \ll & |\,\partial_\gamma
n_a^{\sigma,eq}\,|
\end{eqnarray}
we get
\begin{equation}
\partial_t n_a^{\sigma,eq}\, +\, (\hat e_a)_\gamma\partial_\gamma
n_a^{\sigma,eq}\, =\, -\,\frac{1}{\tau}\,n_a^{\sigma,neq}
\end{equation}
Since $n_a^{\sigma,eq}$ is an equilibrium distribution function, one has
$\partial_t n_a^{\sigma,eq}\, =\, 0$ and so,
\begin{equation}
n_a^{\sigma,neq}\, =\, -\,\tau(\hat e_a)_\gamma\partial_\gamma n_a^{\sigma,
eq}
\end{equation}
Consequently, we get:
\begin{equation}
\Pi^{eq}_{\alpha\beta}\,=\,\frac{c^2}{D}\,(1\,-\,d_0)\,
\rho\delta_{\alpha\beta}\,+\,\sum_\sigma\,
n^\sigma u^\sigma_\alpha u^\sigma_\beta\, -\,\frac{\tau c^2}{D+2}\,\left[
\delta_{\alpha\beta}\partial_\gamma(\rho u_\gamma)\, +\,\partial_\alpha
(\rho u_\beta)\, +\,\partial_\beta (\rho u_\alpha)\right]
\end{equation}

Because of the second order derivatives, only $n_a^{\sigma,eq}$ has a
relevant contribution to $P_\alpha$ and so,
\begin{equation}
P_\alpha\, =\,\frac{c^2}{2(D+2)}\,\nabla^2(\rho u_\alpha)
\end{equation}

The last term in the momentum conservation equation (\ref{eqm}) is obtained
 after a series expansion:
\begin{eqnarray}
\partial_\alpha W & = & \sum_{\sigma ,\bar\sigma ,a}\,G_{\sigma\bar\sigma a}
n_a^\sigma\left[\, n^{\bar\sigma}\, +\,(\hat e_a)_\beta\partial_\beta
 n^{\bar\sigma}
\,\right](\hat e_a)_\alpha\, =\nonumber\\
& = & \sum_{\sigma ,\bar\sigma ,a} G_{\sigma\bar\sigma a} n^\sigma\partial_
\beta n^{\bar\sigma}(\hat e_a)_\alpha(\hat e_a)_\beta
\end{eqnarray}
Taking into account the expression (\ref{ega}) of $G_{\sigma\bar\sigma a}$,
one has
\begin{eqnarray}\label{edefm}
\left( M_{\sigma\bar\sigma}\right)_{\alpha\beta} \,=\,
\sum_a\,G_{\sigma\bar\sigma a}(\hat e_a)_\alpha (\hat e_a)_\beta & = &
\\
\frac{bc^2}{D}\,(\vec m^\sigma\cdot\vec m^{\bar\sigma})\delta_{\alpha\beta}\,
 -\,\frac{3bc^4}{D(D+2)}\,\left[(\vec m^\sigma\cdot\vec m^{\bar\sigma})
\delta_{\alpha\beta}+(\vec m^\sigma)_\alpha(\vec m^{\bar\sigma})_\beta +
(\vec m^\sigma)_\beta(\vec m^{\bar\sigma})_\alpha\right]\, & \quad  &\nonumber
\end{eqnarray}
where the tensor $(M_{\sigma\bar\sigma})_{\alpha\beta}$ is seen to be symmetric
\begin{equation}
\left( M_{\sigma\bar\sigma}\right)_{\alpha\beta}\, =\,\left( M_{\sigma\bar
\sigma}\right)_{\beta\alpha}
\end{equation}
Consequently, we have
\begin{equation}
\partial_\alpha W\, = \, \sum_{\sigma ,\bar\sigma}\,
(M_{\sigma\bar\sigma})_
{\alpha\beta}n^\sigma\partial_\beta n^{\bar\sigma}\, =\,\frac{1}{2}\,
\sum_{\sigma ,\bar\sigma}\,(M_{\sigma\bar\sigma})_{\alpha\beta}\partial_\beta
(n^\sigma n^{\bar\sigma})
\end{equation}

\end{subsection}
\end{section}
\begin{section}{Simulation of structure formation in magnetic fluids}
\begin{subsection}{Dynamic pictures}
As mentioned in the previous chapter, the lattice Boltzmann automaton was
always started after the lattice nodes were quasi-homogeneously filled with
particles having the mean local density $\bar\rho\, =\, 0.5$ with a small
$1\%$ perturbation. The initial values of $n_a^\sigma(i,j,t=0)$ for the local
densities were established in accordance with Eqs.(\ref{e319}) and (\ref{e322})
for each value of the colloidal particle concentration $\phi$ and the field
parameter $h$. For convenience, all runs related to structure formation,
which are described in this section, were performed with the magnetic field
oriented along the $x$ direction.

After a certain number $t$ of iterations, the values of the local particle
 density
\begin{equation}
\rho(i,j,t)\, =\,\sum_{\sigma=0}^S\,\sum_{a=0}^b\, n_a^\sigma(i,j,t)
\end{equation}
and the $x$ component of the magnetisation
\begin{equation}
M_x(i,j,t)\, =\,\sum_{\sigma=1}^S\,\cos\left[\frac{2\pi(\sigma-1)}{S}\right]\,
\sum_{a=0}^b\, n_a^\sigma(i,j,t)
\end{equation}
were retained. The value of the $y$ component of the magnetisation
was always found to be very small ($M_y(i,,j,t)\,\simeq\, 0$) during these
computer runs, a fact which is a direct consequence of the $x$ orientation
of the magnetic field. The mean value of $M_x(i,j,t)$
\begin{equation}
\bar M_x(t)\, =\,<\,M_x(t)\,>\, =\,\frac{1}{(ndx+1)(ndy+1)}\,\sum_{i=0}^{ndx}\,
\sum_{j=0}^{ndy}\, M_x(i,j,t)
\end{equation}
was found to be constant during the time evolution of the automaton, as
expected.

Figure 1 reproduces the automaton state after $t\,=\, 0,\, 10,\, 100$
, $500$, $1000$ and $5000$ iterations, respectively, for $\bar\rho\, =\, 0.5,\,
\phi\, =\, 0.20$ and $h\, =\, 0.5$ ($ndx=ndy=127$). The white points in this
figure have
$M_x(i,j,t)\,>\,\bar M_x(t)\, +\,0.1\cdot\bar M_x(t)$, the gray ones have
$\bar M_x(t)\, <\,M_x(i,j,t)\, \leq\,\bar M_x(t)\, +\,0.1\cdot\bar M_x(t)$,
while the black ones have $M_x(i,j,t)\,\leq\,\bar M_x(t)$. The phase
 separation, i.e., the onset of thread-like clusters orientated along the
 magnetic field direction, is evident. Therefore, the white points in Figure
1 belong
to the high magnetisation phase, while the black ones in the same figure belong
to the low magnetisation phase.

 We have computed separately the mean values of
the $x$ component of the magnetisation in each phase, i.e.,
\begin{eqnarray}
M_x^{high}(t) & = & \frac{1}{N_{high}}\,\sum_{M_x(i,j,t)>\bar M_x(t)}
M_x(i,j,t)\\
M_x^{low}(t) & = & \frac{1}{N_{low}}\,\sum_{M_x(i,j,t)\leq\bar M_x(t)}
M_x(i,j,t)
\end{eqnarray}
where $N_{high}$ and $N_{low}$ are the total numbers of sites belonging to
the high and low magnetisation phase, respectively. A similar procedure was
 adopted also for the evaluation of the mean densities $\rho^{high}(t)$ and
$\rho^{low}(t)$ in the high density and the low density phases, respectively.
The resulting values for the run in Figure 1 are reproduced in
 Table 1.

As the mean values for this run were $\bar\rho(t)\, =\, 0.5$ and $\bar M_x(t)\,
=\, 0.02426$, one can see that there is an intense separation between the
high and low magnetisation phases, while the relative density variations
are much smaller (approx. $2\%$ of the corresponding mean value).

Figure 2 reproduces the same corresponding states as in Figure 1,
obtained for a lower value of the colloidal particle concentration ($\phi\, =
\, 0.05$), while  the other parameters were the same. Although a certain
ordering disposition along the magnetic field direction $x$ is suggested from
this figure, the corresponding high and low magnetisation or density
 values reproduced in Table 2 display no relevant differences. Consequently, no
phase transition occured in this case.

Figures 1 - 2 demonstrate that the lattice Boltzmann model with
dipole-dipole particle interactions can exhibit thermodynamic phase
transitions under certain values of the system parameters $\bar\rho,\,\phi$ and
$h$. The onset of structured patterns which are characteristic to magnetic
fluids, as well as to other systems which undergo spinodal decomposition
 \cite{b3} e.g.,
precipitation-type hard magnetic alloys (AlNiCo8) or type II superconductors,
is well evidentiated in the frame of this model.
\end{subsection}

\begin{subsection}{Phase transitions}
Although the problem of the derivation of an equation of state for our Lattice
Boltzmann model was not considered here, we made some attempts (computer
experiments) in order to evidentiate those values of $\phi$ and $h$ at $\bar
\rho=0.5$ which are characteristic for the phase diagram. The results of
systematic searches performed up to date are displayed in Figures 3 - 5,
where the field dependence of $\bar M_x,\,M_x^{high}$ and $M_x^{low}$ after
$t=5000$ iterations was plotted for $\phi=0.13,\,\phi=0.14$ and $\phi=0.15$,
respectively. The upper and lower magnetisation values are initially close to
the mean value $\bar M_x$ for lower values of the field parameter $h$. As
the field parameter increases, the phase separation is achieved. This process
is clearly seen as the bifurcation of the magnetisation curves in Figures
3 - 5. Consequently, when the magnetic field becomes greater than a
 critical value $h_c\,\equiv\, h_c(\phi)$, the high and low magnetisation
values become clearly different. The critical field values corresponding to
the concentrations in Figures 3 and 4 are $h_c(\phi=0.13)\,\simeq\, 1.1$
and $h_c(\phi=0.14)\,\simeq\, 0.6$. For $\phi=0.15$, the phase separate
even at very low field intensity, $0.0<h_c(\phi=0.15)<0.1$, as one can see
in Figure 5.

The separation of magnetic phases at high values of the magnetic field
intensity is a characteristic process for magnetic fluids \cite{b3,b12,b17}.
This
would be unconvenient for many industrial applications, e.g., high speed
magnetic fluids rotary seals, but this process is partially overcome by the
surfactant layer of the colloidal particles, which always introduces a
supplementary repulsive interaction. Therefore, a more realistic Lattice
Boltzmann model for magnetic fluids should take this aspect into consideration,
in a similar way as in the Monte Carlo models already developed
\cite{b12,b13,b14,b15}.
\end{subsection}
\end{section}

\begin{section}{Simulation of sound propagation in magnetic fluids}
\begin{subsection}{Theoretical background}
We consider a weak perturbation $\rho ',\,\vec u'$ of the equilibrium
solution $\rho^{eq},\,\vec u^{eq}=0$ of the lattice Boltzmann automaton
\cite{bprep}, which satisfies
\begin{eqnarray}
|\,\rho '\, | & \ll & \rho^{eq}\nonumber\\
|\,\vec u'\, | & \ll & 1
\end{eqnarray}
Therefore, we substitute
\begin{eqnarray}
\rho & = & \rho^{eq}\, +\,\rho'\nonumber\\
\vec u & = & 0\, +\,\vec u'
\end{eqnarray}
in the mass and momentum conservation laws and get ($\vec {u'}^\sigma =
\vec u'$), after taking into account that $\rho^{eq}$ is an equilibrium
solution and retaining all terms up to the third order :
\begin{equation}\label{emass1}
\partial_t\rho'\, +\,\rho^{eq}\partial_\alpha u'_\alpha\, +\,\partial_\alpha
(\rho 'u'_\alpha)\, =\, 0
\end{equation}
\begin{eqnarray}\label{emome1}
 & \rho^{eq}\partial_t u'_\alpha\, +\,\partial_t(\rho '\vec u'_a)
\, +\,\frac{c^2}{D}(1-d_0)
\delta_{\alpha\beta}\partial_\beta\rho'\, +\,\rho^{eq}\phi\,
\sum_\sigma\, f^\sigma (\partial_\beta u'_\alpha u'_\beta)\,- & \nonumber\\
 & -\,\frac{\tau c^2}{D+2}\,\left[\delta_{\alpha\beta}\rho^{eq}\partial_\beta
\partial_\gamma u'\, +\,\rho^{eq}\partial_\alpha\partial_\beta u'_\beta\, +\,
\rho^{eq}\partial_\beta\partial_\beta u'_\alpha\right]\,+ & \nonumber\\
 & \frac{c^2}{2(D+2)}\,\rho^{eq}\partial_\beta\partial_\beta u'_\alpha\, = &
\\
 & -\frac{1}{2}\,(2\rho^{eq}\partial_\beta\rho '+2\rho '\partial_\beta\rho ')
\phi^2\,\sum_{\sigma ,\bar\sigma}\, f^\sigma f^{\bar\sigma}\cdot & \nonumber\\
 & \left[\frac{bc^2}{D}(\vec m^\sigma\cdot\vec m^
{\bar\sigma})\delta_{\alpha\beta}\, -\,\frac{3bc^4}{D(D+2)}\,\left[
(\vec m^\sigma\cdot\vec m^{\bar\sigma})\delta_{\alpha\beta}\, +\,(\vec
m^\sigma)_\alpha(\vec m^{\bar\sigma})_\beta\, +\,(\vec m^\sigma)_\beta(\vec
m^{\bar\sigma})_\alpha\,\right]\right] & \nonumber
\end{eqnarray}

The squared sound velocity $(c_S^{1})^2$ is, in the first order approximation,
the coefficient of $\delta_{\alpha\beta}
(\partial_\beta\rho')$:
\begin{equation}\label{cs2}
(c_S^1)^2\, =\,\frac{c^2}{D}\,(1-d_0)\, +\,\rho^{eq}\phi^2\,\left[\,
\frac{bc^2}{D}
\, -\,\frac{3bc^4}{D(D+2)}\,\right]\,\sum_{\sigma ,\bar\sigma}\,f^\sigma
f^{\bar\sigma}(\vec m^\sigma\cdot\vec m^{\bar\sigma})
\end{equation}
This result is a generalisation of the squared sound velocity expression in
current $2D$ Lattice Boltzmann models \cite{b24,b25} and incorporates also the
influence of the magnetic field through the distribution functions $f^\sigma$,
 as well as the influence of the colloidal particle
concentration $\phi$ \cite{x2,bprep}

The sound propagation equation is obtained from the conservation equations
(\ref{emass1}) and (\ref{emome1}), substracting them after their multiplication
with $\partial_t$ and $\partial_\alpha$, respectively, and
taking into account the first-order approximation in the continuity equation:

\begin{eqnarray}\label{eqsf}
\partial_t^2\rho'\, -\,(c_S^1)^2\nabla^2\rho'\, +\,
\frac{(6\tau-1)c^2}{2(D+2)}\,\partial_t(\nabla^2\rho') & = &\nonumber\\
\rho^{eq}\phi\,\sum_{\sigma}\,f^\sigma\partial_\alpha\partial_\beta(u'_\alpha
u'_\beta)\, -\,\rho^{eq}\phi^2(\partial_\alpha\partial_\beta\rho')\,
\sum_{\sigma ,\bar\sigma}\,f^\sigma f^{\bar\sigma}\,(S_{\sigma\bar\sigma})
_{\alpha\beta} & + &\nonumber\\
\partial_\alpha(\rho'\partial_\beta\rho')\phi^2\,\sum_{\sigma ,\bar\sigma}\,
f^\sigma f^{\bar\sigma}\,(M_{\sigma\bar\sigma})_{\alpha\beta} & &
\end{eqnarray}

where
\begin{equation}
(S_{\sigma\bar\sigma})_{\alpha\beta}\, =\,\frac{3bc^4}{D(D+2)}\,\left[
(\vec m^\sigma)_\alpha (\vec m^{\bar\sigma})_\beta\, +\, (\vec m^\sigma)_
\beta (\vec m^{\bar\sigma})_\alpha\,\right]
\end{equation}
and $(M_{\sigma\bar\sigma})_{\alpha\beta}$ was already defined in
Eq.\ref{edefm}.
The supplementary right hand side is responsible for the joint colloidal
particle concentration and magnetic field action. This equation reduces to the
usual damped sound equation when $\phi=0$, i.e., no colloidal magnetic
particles are present.
\end{subsection}
\begin{subsection}{Simulation procedure}
In order to take advantage of the periodic boundaries in our Lattice Boltzmann
computer program, we considered the problem of standing waves in the $x$
direction. Therefore, the lattice was initialized with a cosine perturbation
in the $x$ direction, having the amplitude $\rho_0=0.1$:
\begin{equation}
\rho(x,y,t=0)\, =\, \bar\rho\,\left[ \, 1.\, +\,\rho_0\cos(kx)\, \right]
\end{equation}
i.e.,
\begin{equation}
\rho(i,j,t=0)\, =\, \bar\rho\,\left[\, 1.\, +\,\rho_0\cos(2\pi i/ndx)\,\right]
\end{equation}
Consequently,
\begin{eqnarray}
n^\sigma(i,j,t=0) & = & \rho(i,j,t=0)\phi f^\sigma(\vec H)\nonumber\\
n^0(i,j,t=0) & = & \rho(i,j,t=0)\cdot(1-\phi)
\end{eqnarray}
The time evolution of the lattice automaton was registered over
$n_{iter}\,=\,5000$ iterations, for different values of the
concentration $\phi$ and the field intensity parameter $h$. The
field direction was usually oriented along the $x$ or $y$ axes but, in order
to study the
anisotropy of sound attenuation, the general case when the angle between the
field vector $\vec H$ and the $x$ axis is $\theta$, was also considered.

The general behaviour of the space and time dependence of the perturbation
(after a mediation over the $y$ direction)
\begin{equation}
\rho'(i,t)\, =\,\frac{1}{1+ndy}\,\sum_{j=0}^{ndy}\,\left[\,\rho(i,j,t)-\bar\rho
\,\right]
\end{equation}
was found to be close to the the expression of the attenuated standing waves
with the wavenumber $k=2\pi /ndx$
\begin{equation}
\rho'(x,t)\, =\,\rho_0\,e^{-\alpha t}\cos(kx)\cos(\omega_S t)
\end{equation}
i.e.,
\begin{equation}
\rho'(i,t)\, =\,\rho_0\,e^{-\alpha t}\cos(ki)\cos(\omega_S t)
\end{equation}
for $i=0,\ldots ,ndx$ and $t=0,\ldots ,n_{iter}$.

In order to get the interesting quantities $\alpha$ and $\omega_S=kc_S$,
which are always accesible to experimental measurements, the $x$
dependence was eliminated ($L=2\pi /k$ is the lattice length):
\begin{equation}
a(t)\, =\,\frac{2}{L}\,\int_0^L\,\rho'(x,t) dt\, =\,\rho'_0\,e^{-\alpha t}
\cos(\omega_S t)
\end{equation}
i.e.,
\begin{equation}\label{ampli}
a(t)\, =\,\frac{2}{1+ndx}\,\sum_{i=0}^{i=ndx}\,\rho'(i,t)\,
=\,\rho'_0\,e^{-\alpha t}
\cos(\omega_S t)
\end{equation}
The attenuation coefficient was always found after a least-squares fit of the
local extrema of $a(t)$, while $\omega_S$ was found as the extremum point in
the Fourier spectrum
\begin{equation}
F(\omega)\, =\,\int_0^\infty\,a(t)cos(\omega t)dt \approx\,
\sum_{t=0}^{niter}\,a(t)
\end{equation}
The attenuation coefficient $\alpha$ and the angular frequency $\omega_S$
(or the sound velocity $c_S$, respectively), determined according to the above
mentioned procedure were compared with real experimental data existing in the
literature. The results are discussed below.
\end{subsection}
\begin{subsection}{Sound velocity}
The typical time evolution of the computed
local density perturbation $a(t)$ is
reproduced in Figure 6, for $\phi=0.20$, $h=0.8$ and two perpendicular
orientations of
the magnetic field ($x$ and $y$). The different oscillating frequencies,
due to the anisotropy of sound velocity, and
also the different attenuation of the sound intensity are very clear. The
computed corresponding Fourier spectrum reproduced in Figure 7
also illustrates the sound velocity anisotropy. From
these figures, it is very
clear that, for the same field intensity, the sound velocity is greater
when propagating in the $x$ direction.

On the basis of these first results, a systematic exploration was made
in order to see the influence of the magnetic field intensity at a
fixed concentration $\phi=0.20$. The Fourier spectra
demonstrated that the sound velocity $c_S$ and consequently, also
the angular frequency $\omega_S$, increase in both $x$ and $y$
directions when increasing the value of the
parameter $h$, i.e, when increasing the field intensity while temperature
is maintained constant. As mentioned above, the velocity increase in the $x$
direction is always greater than the corresponding increase in the $y$
 direction.

Figure 8 shows the concentration dependence of the squared angular frequency
$\omega_s^2$, which was obtained after performing computer runs with the
magnetic field $h=0.8$ in the $x$ direction. One can see that the squared
angular frequency
is increasing when increasing concentration, a fact which also agrees
 qualitatively with
the theoretical formula (\ref{cs2}).

The
general behaviour of the sound velocity vs. field intensity, which are
retained by our Lattice Boltzmann computer experiments, i.e., the initial
increase of the velocity, followed by saturation, agrees well
with real experimental measurements e.g., those performed on water based
magnetic fluids \cite{b33}.

\end{subsection}
\begin{subsection}{Sound attenuation}
A systematic investigation of the dependence of the attenuation coefficient
$\alpha$
vs. the angle $\theta$ between the field vector $\vec H$ and the $x$ axis was
carried up. The results are reproduced in Figures
9 $(\phi=0.10,\,h=1.0)$ and 10 $(\phi=0.20,\,
h=0.5)$, respectively . One can see again that the attenuation coefficient is
always greater when the field is oriented along the $x$ direction. Moreover,
when $\theta$ is varied from $0^o$ to $90^o$, the attenuation coefficient
for $\phi=0.20,\,h=0.5$ goes succesively through a maximum and a minimum value.
This feature was already observed in experimental measurements \cite{b34,b35}.
 These results are well explained
by a cluster formation model \cite{b36}. The qualitative agrement
between our computed results and the experimental ones \cite{b35,x4} is
remarkable.
\end{subsection}
\end{section}

\begin{section}{Conclusions}
A lattice Boltzmann model with interacting particles was developed in order to
 simulate the magneto-rheological characteristics of magnetic fluids. In the
frame of this model, $6\, +\,1$ species of particles are allowed to move across
a $2D$ triangular lattice. Among these species, $6$ of them carry an
individual magnetic dipole moment which becomes unchanged during the time
 evolution of
the automaton. These particles interact themselves not only as a result of
local collisions, as in usual Lattice Boltzmann models, but also as a result of
nearest neighbours magnetic dipole-dipole interaction. The relative
distribution of the individual magnetic moments is determined by the intensity
 of an
external static magnetic field acting on the whole system.
This model exhibits some relevant characteristics of real magnetic fluids,
 i.e.,
structure formation as a result of magnetic field induced gas-liquid
phase transition and anisotropy of these structures. The magnetic field induced
anisotropy of sound velocity and attenuation in magnetic fluids is also well
evidentiated in the frame of this model.

The extension of this model in order to allow the particle system to be
subjected to time variations of the applied magnetic field amplitude and/or
 orientation through the introduction of a second relaxation time which
may take into acccount the orientation relaxation of the magnetic colloidal
particles, may serve as a basis for the analysis of hot topics related to the
 magneto - rheological
behaviour of magnetic fluids (surface instabilities, pipe flow, magnetic
 B\'enard convection,
Taylor vortices formation, heat transfer), as well as an efficient approach
to the simulation of the behaviour of some magnetic fluid industrial devices,
 such
as rotary seals, dampers and inductive transducers, onset of the rotary motion
 of magnetic fluids under the action of
rotating magnetic fields and phase transitions induced in magnetic fluids
under the action of transient magnetic fields.
\end{section}


\acknowledgments
{This work was done in the frame of a two months research stage
(August - September 1993) financed by the Deutscher Akademischer
Austauschdienst (DAAD), Kennedyallee 50, D-53175 Bonn, Germany, which is
gratefully acknowledged.

The author is indebted to Professor John ARGYRIS, director of the
Institute for Computer Applications (ICA), University of Stuttgart,
Pfaffenwaldring 27, D-70569 Stuttgart, Germany, for his kind hospitality at
 this
institute during the DAAD research stage. The excellent working
conditions
and computer facilities existing at ICA are particularly
appreciated.

The author is also indebted to Dipl.-Ing. Gerald P\"ATZOLD for
fruitful discussions and valuable computer assistance.
}

\newpage
\begin{quasitable}\label{tab1}
\begin{center}
\begin{tabular}{rcccc}
 & & & & TABLE 1.\medskip\\
\hline\\
t & $\rho^{high}$ & $\rho^{low}$ & $M_x^{high}$ & $M_x^{low}$\medskip\\
\hline\\
0 & 0.50000 & 0.50000 & 0.02431 & 0.02418\\
10 & 0.50000 & 0.49999 & 0.02427 & 0.02423\\
100 & 0.50003 & 0.49996 & 0.02435 & 0.02414\\
500 & 0.51433 & 0.49145 & 0.05967 & 0.00147\\
1000 & 0.51276 & 0.49083 & 0.05440 & 0.00070\\
5000 & 0.51298 & 0.49083 & 0.05394 & 0.00124\smallskip\\
\hline
\end{tabular}
\end{center}
\end{quasitable}
\vskip 2cm
\begin{quasitable}\label{tab2}
\begin{center}
\begin{tabular}{rcccc}
& & & & TABLE 2.\medskip\\
\hline\\
t & $\rho^{high}$ & $\rho^{low}$ & $M_x^{high}$ & $M_x^{low}$\medskip\\
\hline\\
0 & 0.50000 & 0.50000 & 0.0060778 & 0.0060475\\
10 & 0.50000 & 0.50000 & 0.0060654 & 0.060599\\
100 & 0.50000 & 0.50000 & 0.0060635 & 0.0060619\\
500 & 0.500000016 & 0.499999985 & 0.0060629 & 0.0060624\\
1000 & 0.500000013 & 0.499999986 & 0.0060628 & 0.0060625\\
5000 & 0.500000017 & 0.499999984 & 0.0060627 & 0.0060626\smallskip\\
\hline
\end{tabular}
\end{center}
\end{quasitable}
\newpage

List of figure captions\bigskip

\begin{enumerate}
\item Dynamic evolution of the local magnetisation after $t\,=\,0,\,10,\,100,\,
500,\,1000$ and $5000$ time steps, for $\bar\rho\,=\,0.5$, $\phi\,=\,0.20$
and $h\,=\,0.5$.
\item Dynamic evolution of the local magnetisation after $t\,=\,0,\,10,\,100,\,
500,\,1000$ and $5000$ time steps, for $\bar\rho\,=\,0.5$, $\phi\,=\,0.05$
and $h\,=\,0.5$.
\item Field parameter $(h)$ dependence of the mean $(\bullet)$, high
 $(\Diamond)$
and low $(\Box)$ magnetisation values after $t\,=\,5000$ time steps, for
$\bar\rho\,=\,0.5$ and $\phi\,=\,0.13$.
\item Field parameter $(h)$ dependence of the mean $(\bullet)$, high
 $(\Diamond)$
and low $(\Box)$ magnetisation values after $t\,=\,5000$ time steps, for
$\bar\rho\,=\,0.5$ and $\phi\,=\,0.14$.
\item Field parameter $(h)$ dependence of the mean $(\bullet)$, high
 $(\Diamond)$
and low $(\Box)$ magnetisation values after $t\,=\,5000$ time steps, for
$\bar\rho\,=\,0.5$ and $\phi\,=\,0.15$.
\item Typical time evolution of the computed local density perturbation $a(t)$
for $\phi\,=\,0.20$, $h\,=\,0.8$ and two orientations of the magnetic field
($x$ and $y$).
\item Computed Fourier spectrum for the two curves in Figure 6.
\item Concentration dependence of the squared angular frequency $\omega_S^2$,
obtained for $h\,=\,0.8$.
\item Dependence of the attenuation coefficient $\alpha$ vs. the angle $\theta$
for $\phi\,=\,0.10$ and $h\,=\,1.0$.
\item Dependence of the attenuation coefficient $\alpha$ vs. the angle $\theta$
for $\phi\,=\,0.20$ and $h\,=\,0.5$.

\end{enumerate}

\end{document}